\documentclass[showpacs,pra,twocolumn]{revtex4-2}

\usepackage[ruled]{algorithm2e}
\usepackage{verbatim}
\usepackage{epsfig}
\usepackage{graphicx}
\usepackage{subfigure}
\usepackage{amsmath}
\usepackage{dcolumn}
\usepackage{stmaryrd}
\usepackage{mathrsfs}
\usepackage{amsthm}
\usepackage{amssymb}
\usepackage{pifont}
\usepackage{bm}
\usepackage{latexsym}
\usepackage{hyperref}
\usepackage{color}
\usepackage{xcolor}

\begin{document}
	\title
		{Stochastic learning control of adiabatic speedup in a non-Markovian open qutrit system}
	\author
		{
		Yang-Yang Xie\textsuperscript{1}, Feng-Hua Ren\textsuperscript{2}, Run-Hong He\textsuperscript{1}, Arapat Ablimit\textsuperscript{1}, Zhao-Ming Wang\textsuperscript{1}
		}
	\altaffiliation
		{Corresponding author: wangzhaoming@ouc.edu.cn}
	\affiliation
		{
		\textsuperscript{1}College of Physics and Optoelectronic Engineering, Ocean University of China, Qingdao 266100, China \\
		\textsuperscript{2}School of Information and Control Engineering, Qingdao University of Technology, Qingdao 266520, China
		}
	\date{\today}

\begin{abstract}
	Precise and efficient control of quantum systems is essential to perform quantum information processing tasks. In terms of adiabatic speedup via leakage elimination operator approach, for a closed system, the ideal pulse control conditions have been theoretically derived by P-Q partitioning technique. However, it is a challenge to design the corresponding control pulses for an open system, which requires to tackle noisy environments. In this paper, we apply the stochastic search procedures to an open qutrit system and successfully obtain the optimal control pulses for significant adiabatic speedup. The calculation results show that these optimal pulses allow us to acquire higher fidelities than the ideal pulses. The improvement of fidelity is large for relatively strong system-bath coupling strength and high bath temperature. For certain coupling strength and bath temperature, the maximal improvement can be achieved for a critical characteristic frequency which represents the memory time of the environment. Our investigation indicates that the stochastic search procedures are powerful tools to design control pulses for combating the detrimental effects of the environment.
\end{abstract}

\maketitle

\section{Introduction}
 	Robust and accurate control of quantum systems is of paramount importance in the field of quantum computation and quantum information processing, such as adiabatic quantum computing \cite{adiabaticqc1,adiabaticqc2,adiabaticqc3}, adiabatic quantum state transmission \cite{chenyifei pla paper,statetransmission1,PhysRevB.76.094411,PhysRevB.91.134303}, or quantum gates \cite{qgates}. In general, quantum control is employed to find strategies of quantum state evolution from an initial state towards a specified target state. The design of such strategies has been widely studied both theoretically and numerically. For theory, it includes Lyapunov quantum control \cite{Lyapunov1,Lyapunov2}, geometric control \cite{Geometric} and the Pontryagin maximum principle \cite{Pontryagin}, etc. For more complicated systems, numerical algorithms are developed, like stochastic gradient descent (SGD) and Adam algorithms \cite{sgd2}, Krotov algorithm \cite{krotov1,krotov2,krotov3,krotov4,krotov5}, gradient ascent pulse engineering algorithm \cite{grape1,grape2}, chopped random basis algorithm \cite{crab1,crab2} and distributed proximal policy optimization algorithm \cite{dppo1,dppo2}.
 
 	In adiabatic quantum computation \cite{2001Robustness,Mandr2015Adiabatic,Pyshkin2016Expedited}, the system needs to evolve along an adiabatic path and normally an infinitely long evolution time is necessary \cite{wang2018}. However, undesirable transitions between various eigenstates of the system \cite{pulsecondition,wangrui} will occur for a short evolution time, which is required in performing actual tasks. Adiabatic speedup, or shortcut to adiabaticity, has been put forward to adiabatically accelerate the evolution process to restrain the transitions, including transitionless quantum driving \cite{transitionless1,transitionless2}, invariant-based inverse engineering \cite{invariant-based1,invariant-based2,invariant-based3}, acceleration of the adiabatic passage \cite{adiapassageacce1,adiapassageacce2}, superadiabatic driving \cite{superadiadrivin1,superadiadrivin2,superadiadrivin3}, counterdiabatic driving \cite{cd1,cd2}, fast-forward approach \cite{fastforward1,fastforward2}. Moreover, in reality the system will be inevitably embedded in its surrounding environment. The system-environment interaction will also destroy the adiabaticity. This destructiveness will accumulate and become severer over time, which incorporates dissipation, decoherence and other effects \cite{wang2018}. Adiabatic speedup in open systems has also been proposed. One scheme is to use the leakage elimination operator (LEO) approach, which has been studied in closed or open systems \cite{leo1,leo2,leo3,wang2018}. The LEO Hamiltonian can be realized by a sequence of control pulses \cite{ren2017,wangrui,pulsecondition}. The pulse control conditions have been derived theoretically by P-Q partitioning technique. However, these control conditions can only be deduced in closed systems. For a  weak environment, the system can be considered nearly closed and the ideal closed-system pulses will function well in this case \cite{wang2018,wang2020,wujing}. While for a relatively strong environment, pulse control conditions are not as effective as in the closed case. In this case, it is necessary to take into account the environmental effects for the pulse design \cite{qgates}. Then it is interesting to investigate whether or not in open systems, stochastic search procedures (SSPs) are able to search the optimal pulses directly and correct the fidelity decrease due to the environment effects.
 	
 	Open system dynamics have been extensively investigated in the past two decades \cite{markovapprox1,langevin1}. When the memory effects of the environment are neglected, Markovian approximation \cite{markovapprox2,markovapprox3} is often used to study the evolution of the system, e.g. the Langevin equations \cite{langevin2} or the master equations \cite{langevin1}. However, when the memory effects have to be taken into account, a non-Markovian description is required. For example, a recent experiment indicates that the bath coupled to an optomechanical system is non-Markovian \cite{Groblacher}. With the experimental technique on environment engineering, it might be possible to observe the non-Markovian dynamics of open quantum systems \cite{Liu}. Normally it is a daunting task to solve the non-Markovian dynamics of the open systems. The quantum state diffusion (QSD) equation approach provides a very promising technique \cite{qsde4,qsde6}. In this paper, we adopt QSD approach to solve the dynamics of a three-level system and combine it with quantum optimal control based on the SSPs \cite{sgd2} to find the optimal control pulses for adiabatic speedup in a non-Markovian environment. To be specific, the model we consider is a qutrit linearly coupled to a finite-temperature heat bath. Compared with the ideal closed-system pulses, the fidelity improvement $Im$ under different region of parameters is discussed. Our results show that significant $Im$ can be obtained for relatively strong system-bath coupling strength $\Gamma$ and high bath temperature $T$. However, as the effects of the control become weak in a more Markovian bath, there is a critical parameter $\gamma$ which corresponds to the maximum of $Im$. We find that the optimal open-system pulses designed by the SSPs bear the advantage of correcting the fidelity decrease induced by the environment.

\section{Model and method}
\subsection{The model and the Hamiltonian}
	Suppose a quantum system is immersed in a multimode bosonic bath, the total Hamiltonian $H_{tot}$ then consists of three parts:
		\begin{equation}
			H_{tot}=H_{s}+H_{b}+H_{int}.
		\end{equation}
	Here $H_{s}$ is merely the system Hamiltonian and
		\begin{equation}
			H_{b}=\sum_{k}\omega_{k}b_{k}^{\dag}b_{k} 
		\end{equation}
	is the Hamiltonian of the bosonic bath (setting $\hbar=1$), with $\omega_{k}$ indicating the frequency of the $k$th mode of the bath and $b_{k}^{\dag}$ ($b_{k}$) standing for the creation (annihilation) operator. Moreover, the system-bath interaction Hamiltonian $H_{int}$ reads
		\begin{equation}
			H_{int}=\sum\limits_{k}(g_{k}^{\ast}L^{\dag}b_{k}+g_{k}Lb_{k}^{\dag}).
		\end{equation}
	Here the Lindblad operator $L$ describes the system-bath coupling and $g_{k}$ denotes the coupling constant between the system and the $k$th mode of the bosonic bath.
	
	Now we adopt the QSD approach \cite{qsde4,qsde6,wang2021,o1} to calculate the system dynamics. Accordingly, the master equation of the open system in a non-Markovian finite-temperature bath can be constructed as
		\begin{eqnarray}
			\frac{\partial}{\partial t}\rho_{s}&=&-i[H_{s},\rho_{s}]+[L,\rho_{s}\overline{O}_{z}^{\dag}(t)]-[L^{\dag},\overline{O}_{z}(t)\rho_{s}]\notag \\
			&&+[L^{\dag},\rho_{s}\overline{O}_{w}^{\dag}(t)]-[L,\overline{O}_{w}(t)\rho_{s}]. 
			\label{master}
		\end{eqnarray}
	Here $\overline{O}_{z,(w)}(t)=\int_{0}^{t}ds{\alpha}_{z,(w)}(t-s){O}_{z,(w)}(t)$ with the environmental correlation functions ${\alpha}_{z,(w)}(t-s)$, and the operators ${O}_{z,(w)}$ are defined by an ansatz (For details see \cite{o1,o2}).

	As in Ref.~\cite{wang2021}, the master equation in Eq.~(\ref{master}) can be numerically solved with the help of the following closed equations
	\begin{equation}
		\frac{\partial \overline{O}_{z}}{\partial t}=(\frac{\Gamma T\gamma}{2}-\frac{i\Gamma\gamma^{2}}{2})L-\gamma\overline{O}_{z}+[-iH_{s}-(L^{\dag }\overline{O}_{z}+L\overline{O}_{w}),\overline{O}_{z}], \label{o1}
	\end{equation}
	\begin{equation}
		\frac{\partial \overline{O}_{w}}{\partial t}=\frac{\Gamma T\gamma}{2}L^{\dag}-\gamma\overline{O}_{w}+[-iH_{s}-(L^{\dag }\overline{O}_{z}+L\overline{O}_{w}),\overline{O}_{w}]. \label{o2}
	\end{equation}
	Here $\Gamma$ represents the strength of the system-bath coupling and $\gamma$ stands for the characteristic frequency of the bath. They both are dimensionless real parameters. Furthermore, the environmental memory time $1/\gamma$ characterizes the memory capacity of the relevant bath. For small $\gamma$, non-Markovian properties can be observed. The larger $\gamma$ is, the bath turns into more Markovian and has less memory capacity due to the shrinking environmental memory time.
	
	In the Markovian limit (i.e. $\gamma \to \infty$), Eqs.~(\ref{o1}) and (\ref{o2}) become $\overline{O}_{z}=\frac{\Gamma T}{2}L$ and $\overline{O}_{w}=\frac{\Gamma T}{2}L^{\dagger}$. The master equation in Eq.~(\ref{master}) therefore reduces to the Lindblad form \cite{wang2021}
		\begin{eqnarray}
			\frac{\partial }{\partial t}\rho_{s}&=&-i[H_{s},\rho_{s}]+\frac{\Gamma T}{2}[(2L\rho_{s}L^{\dagger}-L^{\dagger}L\rho_{s}-\rho_{s}L^{\dagger}L)\notag \\
			&&+(2L^{\dag}\rho_{s}L-LL^{\dag}\rho_{s}-\rho_{s}LL^{\dag})].
           \label{markov}
		\end{eqnarray}

	In this work, we take the qutrit system as an example, the Hamiltonian reads \cite{wang2018,wujing}
		\begin{equation}
			H_{s}(t)=\omega_{0}\left[(1-t/T_{tot})J_{z}+t/T_{tot} J_{x}\right].
		\end{equation}
	Here $J_{z}=\vert 2\rangle \langle 2\vert-\vert 0\rangle \langle 0\vert$ and $J_{x}=(\vert 2\rangle\langle 1\vert+\vert 1\rangle\langle 2\vert+\vert 1\rangle\langle 0\vert+\vert 0\rangle\langle 1\vert)/\sqrt{2}$. $T_{tot}$ is the total evolution time and the Lindblad operator is considered as $L=J_{-}=\sqrt{2}(\vert 0\rangle\langle 1\vert+\vert 1\rangle\langle 2\vert)$ as an example. We set the initial spacing of the two adjacent energy levels $\omega_{0}=1$. With the ground state $\vert 0\rangle$ as our initial state, the dynamical evolution process is expected to end up with the given target state $(\vert 0\rangle-\sqrt{2}\vert 1\rangle+\vert 2\rangle)/\sqrt{2}$. Here we use the fidelity $F(t)=\sqrt{\left\langle E(t)\right\vert \rho _{s}(t)\left\vert E(t)\right\rangle }$ to measure the adiabaticity during the evolution process \cite{wang2018}, where $\rho_{s}(t)$ is the reduced density matrix and $\left\vert E(t)\right\rangle$ is the noiseless instantaneous eigenstate of the system.

\begin{figure}[!htbp]
\centering
\includegraphics[width=\columnwidth]{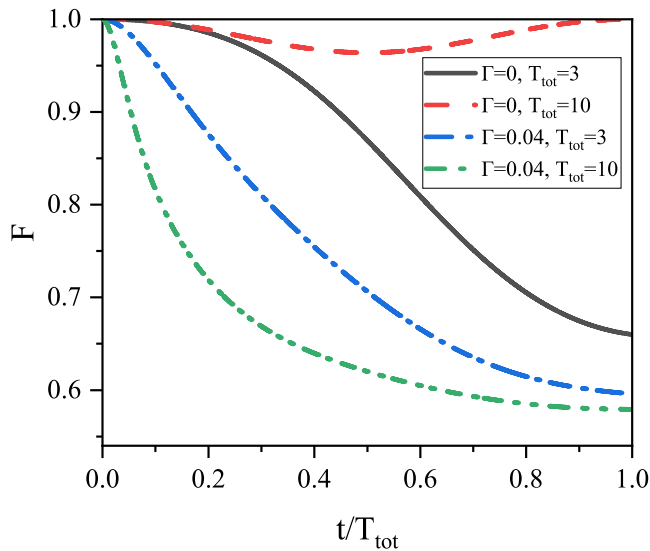}
\caption{(Color on line) The fidelity $F$ versus the rescaled time $t/T_{tot}$ with ($\Gamma=0.04$, $\gamma=2$ and $T=10$) and without ($\Gamma=0$) environment when $T_{tot}=3,10$.}
\label{f1}
\end{figure}

	In Fig.~\ref{f1} we plot the fidelity $F$ versus the rescaled time $t/T_{tot}$ with and without environment for $T_{tot}=3,10$, respectively. For $\Gamma=0$, i.e. closed-system cases, this qutrit system is in an adiabatic regime when $T_{tot}=10$ but in a non-adiabatic regime when $T_{tot}=3$. Notice that for $\Gamma=0.04$, $F(T_{tot}=10)$ is lower than $F(T_{tot}=3)$ due to the effects of the environment over time.

\subsection{Adiabatic speedup under external control}
	 For this model, the fidelity decreases with the ever-growing system-bath coupling strength $\Gamma$, parameter $\gamma$ and temperature $T$ \cite{wujing}. Also, this destructiveness becomes server with the growth of the total evolution time $T_{tot}$. Quantum optimal control schemes have been proposed to realize adiabatic speedup in a non-adiabatic regime by adding an LEO Hamiltonian to the system. As a result, the destructiveness can be reduced. The LEO can be implemented by a sequence of control pulses, which can be constructed as \cite{wang2018,wujing}
		\begin{equation}
			H_{LEO}(t)=c(t)H_{s}(t).
		\end{equation}
	Here $c(t)$ is a control function. Consequently, the modulated Hamiltonian reads
		\begin{equation}
			H_{c}(t)=[1+c(t)]H_{s}(t).
		\end{equation}

	Physically, the control function $c(t)$ can be implemented by a sequence of zero-area pulses, where the integral of the pulse intensity in the time domain is zero in one pulse period \cite{wujing}. To obtain an effective control, the pulse intensity and duration are needed to satisfy certain relations (For details see \cite{pulsecondition}). The control conditions for various types of zero-area pulses have been theoretically derived by P-Q partitioning technique, which can only be deduced in closed systems \cite{chenyifei pla paper,pulsecondition,zhanglicheng}.	For instance, when the energy gap between two adjacent energy levels is constant, for sinusoidal pulses $c(t)=I\sin(\omega t)$ the corresponding pulse control condition is $J_{0}(\frac{I\tau}{\pi})=0$. Here $I$ stands for the amplitude of the pulse intensity and $\tau$ is the half pulse period, $J_{0}(x)$ denoting the zero-order Bessel function of the first kind. However, for the qutrit system the energy gap between the ground state and the first excited state $\Delta E_{10}$ is time-dependent, instead of a constant. In this case the amplitude of the pulse intensity $I$ needs to be tuned \cite{pulsecondition,wujing}:
		\begin{equation}
			I\left ( t \right )=I/\Delta E_{10}\left ( t \right )
			\label{pulse1}
		\end{equation}
	where $\Delta E_{10}(t)=\sqrt{T^{2}_{tot}-2T_{tot}t+2t^{2}}/T_{tot}$. The control functions such as rectangular and triangular ones have also been investigated \cite{wang2018,pulsecondition,zhanglicheng}. 
	
	The pulse control conditions in above theoretical derivation are only applicable in closed systems, and will lose their effectiveness in open systems due to the detrimental effects of the environment \cite{wang2020}. In this work we use the SSPs to directly find the optimal pulses in an open system, which have the advantage that the environmental effects are also taken into account. To compare with the ideal sine pulses, here the control function $c(t)$ is also taken as sinusoidal with a time-dependent pulse intensity $I\left(t \right)$:
		\begin{equation}
			c(t)=I(t)\sin(\omega t). 
			\label{pulse2}
		\end{equation}
	Here $I(t)$ is a $N$ segment piecewise constant function, whose $N$ values are drawn in order from the pulse intensity sequence $\textbf{\textit{I}}=[I_{0},I_{1},\cdots ,I_{N-1}]$ and take the equal time interval $\Delta t=T_{tot}/N$ ($\omega=2\pi/\Delta t$). We set $N=5$ and $T_{tot}=3$ ($T_{tot}=3$ lies in a non-adiabatic regime) throughout this work. The time step size is taken as $T_{tot}/10000$ in our calculation. Note that zero-area pulse conditions \cite{wang2018,wangrui,pulsecondition} are followed in the procedures as in theoretical derivation.

\subsection{Optimal pulse design via SSPs}
	Our goal is to optimize the pulse intensity sequence \textbf{\textit{I}} to design a better control function for significant adiabatic speedup, which allows us to reduce much more effects of the bath than the ideal closed-system control function. This optimization goal is encoded to minimize the loss, or the fidelity error, which is normally defined as
		\begin{equation}
			L(\textbf{\textit{I}})=1-F(\textbf{\textit{I}})+\lambda c_{max}.
			\label{loss}
		\end{equation}
	Here $c_{max}$ is the maximum of the control function $c(t)$. In Eq.~(\ref{loss}), there is a competition between the infidelity $1-F(\textbf{\textit{I}})$ and maximal control strength $c_{max}$ for the calculation of loss $L$. Here we introduce a relaxation parameter $\lambda$ to control their weights \cite{sgd1}. When $\lambda=0$, the SSPs are inclined to reach a minimal infidelity at the cost of ever-growing control strengths, which may not be easy to be realized experimentally. We can modulate $\lambda$ to restrain this tendency. As the attainable final fidelity $F(T_{tot})$ varies with environmental parameters and other settings, the value of $\lambda$ changes accordingly. There is a simple rule that for the same parameters a larger $\lambda$ usually corresponds to a smaller $c_{max}$.
	
	SGD is one of the simplest gradient-based optimization algorithm and can be applied here to construct an iterative process to find satisfactory control pulses and minimize Eq.~(\ref{loss}) \cite{sgd0}. The procedure is presented below.
	\begin{algorithm}
		\caption{SGD}
		\KwData{learning rate $\alpha$ and the initial pulse intensity sequence $\textbf{\textit{I}}^{i}$} 
		\KwResult{the final pulse intensity sequence $\textbf{\textit{I}}^{f}$}
		Set iteration counter $k=0$. \\
		\Repeat
		{$L(\textbf{\textit{I}}^{k})<\xi$ or $k>k_{max}$}
		{Set $k=k+1$. \\
			Compute the gradient $\textbf{\textit{g}}^{k}=\triangledown_{\textbf{\textit{I}}^{k}}L(\textbf{\textit{I}}^{k})$. \\
			Set $\textbf{\textit{I}}^{k}=\textbf{\textit{I}}^{k-1}-\alpha \textbf{\textit{g}}^{k}$. \\}
	\end{algorithm}
	
	However, SGD may converge slowly \cite{sgd1,sgd2}. To speed up this convergence, several improvements have been proposed, among which Adam is an efficient and scalable one \cite{sgd2}. The major distinction between SGD and Adam is that Adam is able to tune the learning rate for each parameter according to the gradient of each iteration. When the gradient is large, the learning rate is modulated to be small and vice versa. Moreover, to estimate the gradient steadily, an exponential moving average is considered on the fly. Then an accelerated and steady convergence is supposed to be acquired. SGD and Adam have been used to sample at each iteration from the distribution of the parameter uncertainty and both algorithms behave well with respect to benchmarks \cite{sgd2}. The specific depiction of Adam is given below.
	\begin{algorithm}
		\caption{Adam}
		\KwData{learning rate $\alpha$, EMA parameters $\beta_{1}$ and $\beta_{2}$, the epsilon $\varepsilon$ and the initial pulse intensity sequence $\textbf{\textit{I}}^{i}$} 
		\KwResult{the final pulse intensity sequence $\textbf{\textit{I}}^{f}$}
		Set iteration counter $k=0$, the first moment vector $\textbf{\textit{m}}^{i}=\textbf{\textit{0}}$ and the second moment vector $\textbf{\textit{v}}^{i}=\textbf{\textit{0}}$. \\
		\Repeat
		{$L(\textbf{\textit{I}}^{k})<\xi$ or $k>k_{max}$}
		{Set $k=k+1$. \\
			Compute the gradient $\textbf{\textit{g}}^{k}=\triangledown_{\textbf{\textit{I}}^{k}}L(\textbf{\textit{I}}^{k})$. \\
			Compute the exponential moving averages $\textbf{\textit{m}}^{k}=\beta_{1}\textbf{\textit{m}}^{k-1}+(1-\beta_{1})\textbf{\textit{g}}^{k}$, $\textbf{\textit{v}}^{k}=\beta_{2}\textbf{\textit{v}}^{k-1}+(1-\beta_{2})(\textbf{\textit{g}}^{k})^{2}$.    \\
			Compute the bias-corrected moment vectors $\hat{\textbf{\textit{m}}^{k}}=\textbf{\textit{m}}^{k}/[1-(\beta_{1})^{k}]$, $\hat{\textbf{\textit{v}}^{k}}=\textbf{\textit{v}}^{k}/[1-(\beta_{2})^{k}]$.               \\
			Set $\textbf{\textit{I}}^{k}=\textbf{\textit{I}}^{k-1}-\alpha \hat{\textbf{\textit{m}}^{k}}/(\sqrt{\hat{\textbf{\textit{v}}^{k}}}+\varepsilon)$.        \\}
	\end{algorithm}

\begin{figure}[!htbp]
\centering
\includegraphics[width=\columnwidth]{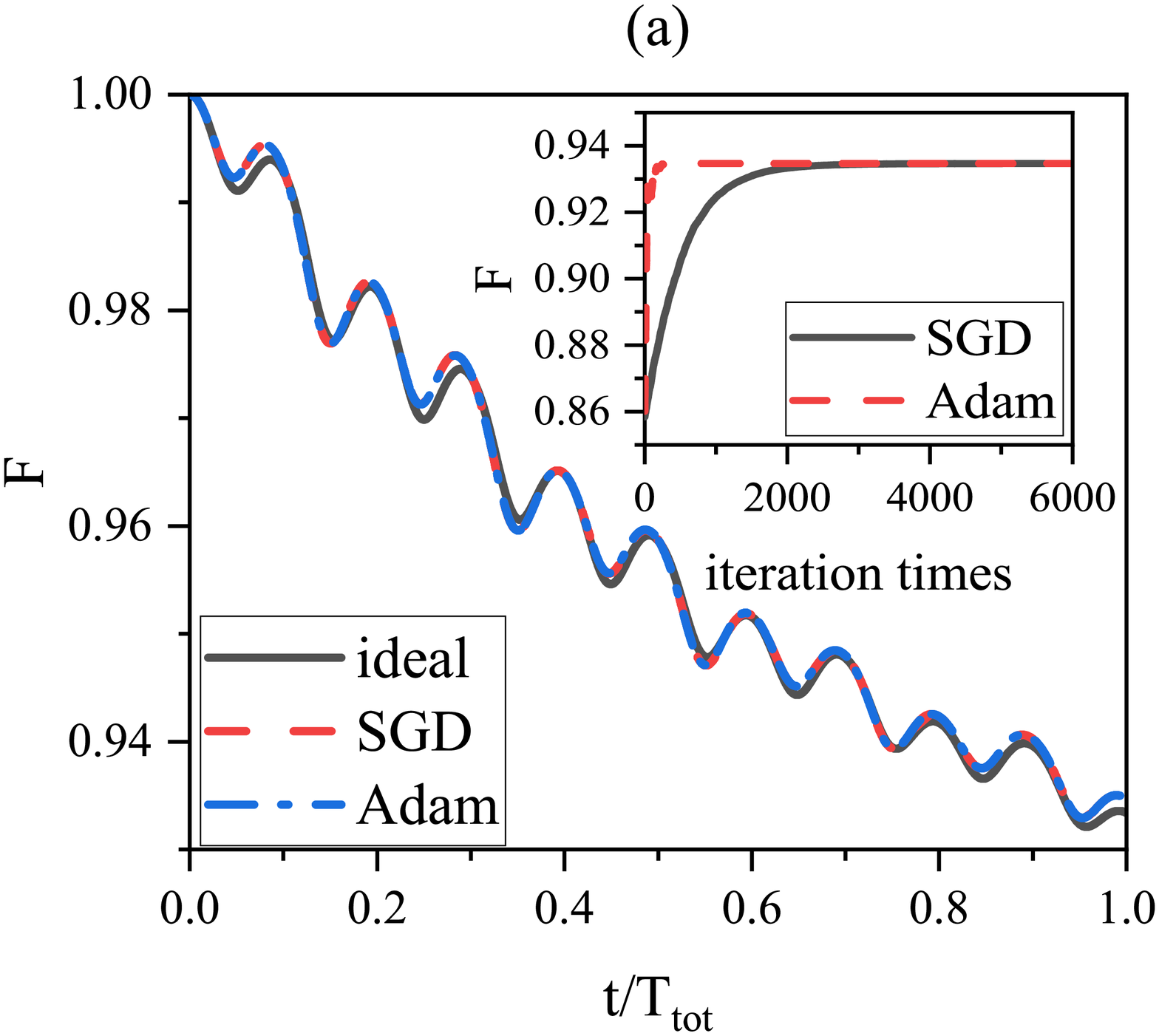}
\includegraphics[width=\columnwidth]{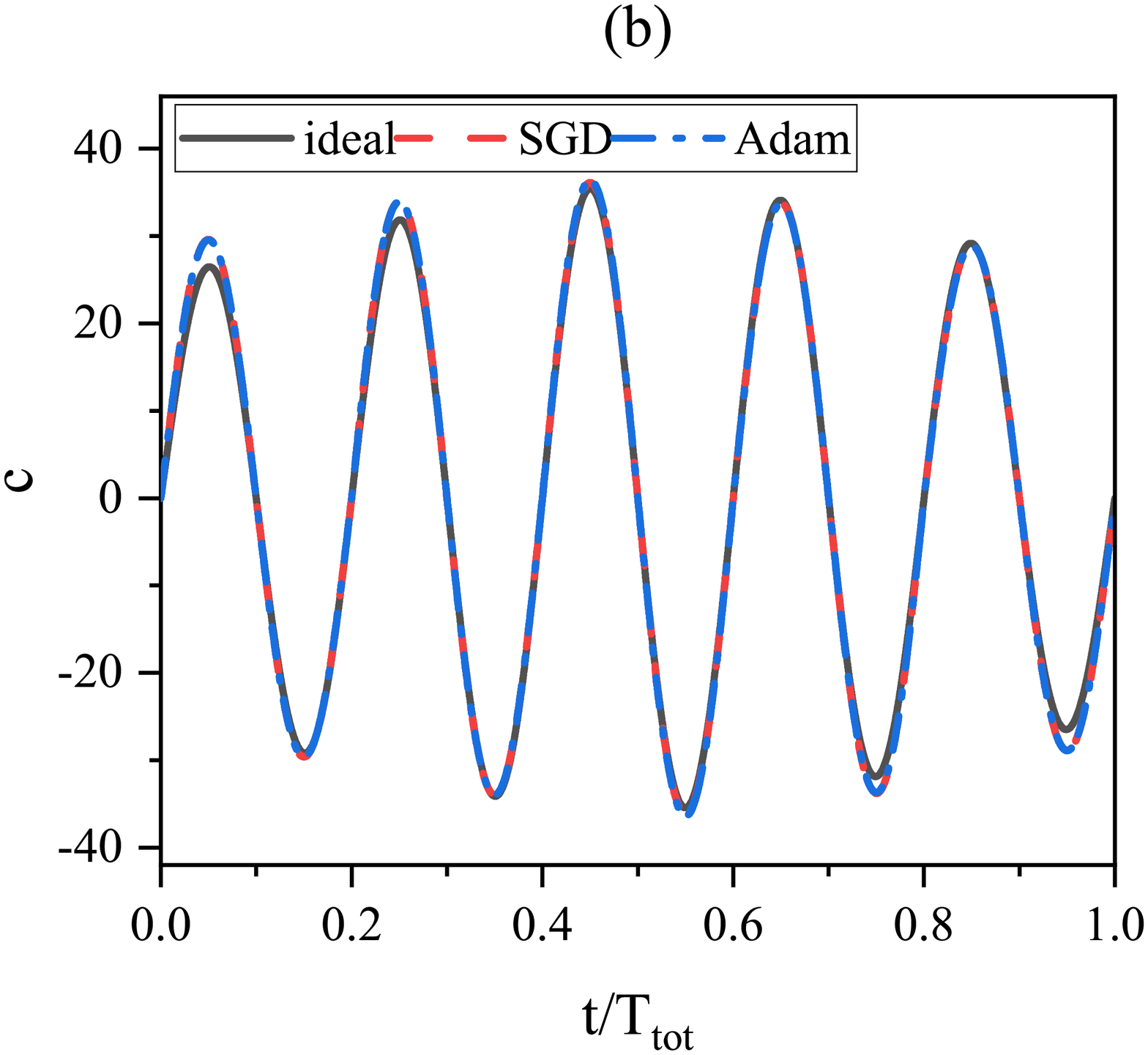}
\caption{(Color on line) (a) The fidelity $F$ versus the rescaled time $t/T_{tot}$ with the help of the ideal closed-system pulses and the optimal open-system pulses (optimized by SGD and Adam). Here $\Gamma=0.04$, $\gamma=4$ and $T=10$. For SGD (Adam), we choose the learning rate $\alpha=10~(1)$. The maximum of iteration times $k_{max}=6000$ for both algorithms. (b) The corresponding profiles of control pulses used in Fig.~\ref{f2}(a).}
\label{f2}
\end{figure}

	Here the initial control $c(t)$ ($I(t)$) is usually constructed by either experience or guess. The iterative process will be terminated if the loss after an iteration $L(\textbf{\textit{I}}^{k})$ is less than the given threshold $\xi$ or the iteration times $k>k_{max}$ (setting $\xi=0.001$).

\section{Results and discussions}
	In this section we take the sinusoidal function as an example, employ these two SSPs (SGD and Adam) to design pulses and compare their performances with the ideal closed-system pulses to demonstrate the adiabatic speedup. The ideal closed-system pulse function is given by Eqs. (\ref{pulse1}) and (\ref{pulse2}) and we take $\textbf{\textit{I}}^{i}=[25.185,25.185,\cdots ,25.185]$ as our initial choice to design optimal open-system pulses, which are different from the ideal pulses. Moreover, in actual experiments, the control intensity can not be infinite. The achievable pulse intensity depends on the physical system and the control agent. In this paper, we limit the control intensity in the range $\left |c(t)\right |<50$.
	
\begin{figure}[!htbp]
\centering
\includegraphics[width=\columnwidth]{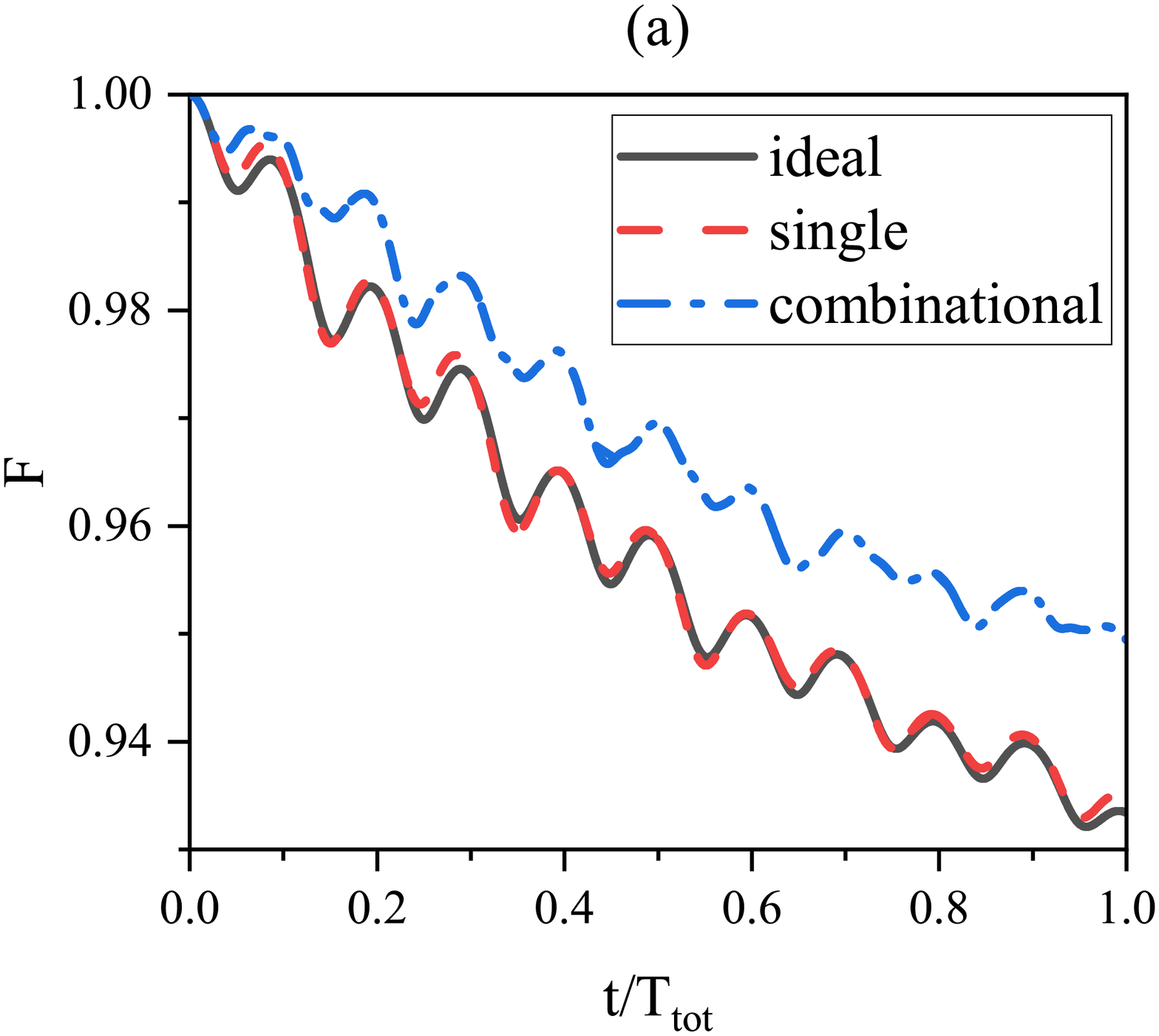}
\includegraphics[width=\columnwidth]{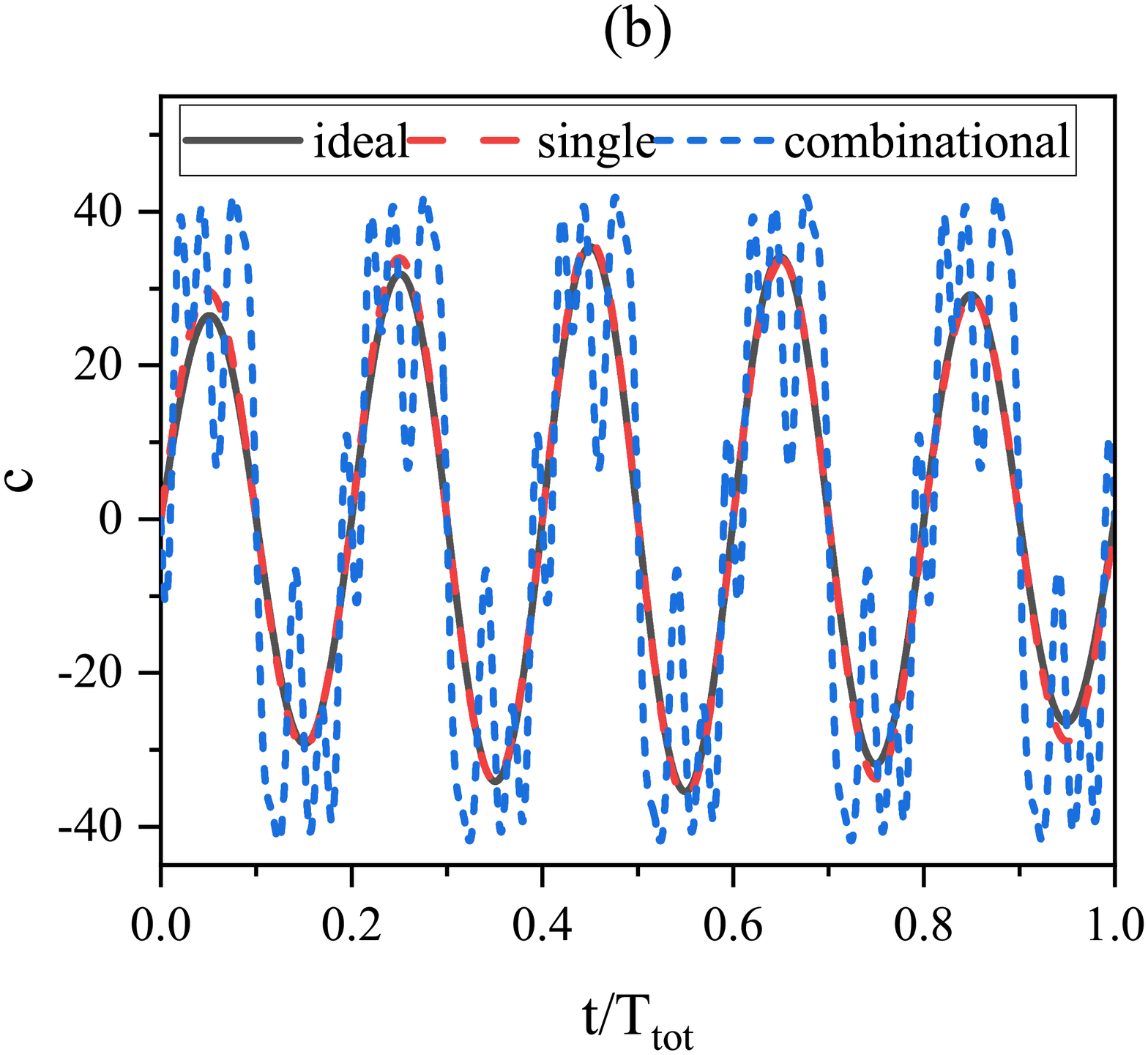}
\caption{(Color on line) (a) The fidelity $F$ versus the rescaled time $t/T_{tot}$ with the help of the ideal closed-system pulses and the two kinds of optimal open-system pulses (optimized by Adam). Here $\Gamma=0.04$, $\gamma=4$ and $T=10$. We choose the learning rate $\alpha=1$ and the maximum of iteration times $k_{max}=1000$. (b) The corresponding profiles of control pulses used in Fig.~\ref{f3}(a).}
\label{f3}
\end{figure}

	Fig.~\ref{f2}(a) plots the fidelity $F$ versus the rescaled time $t/T_{tot}$ under the ideal closed-system and the optimal open-system pulse control, respectively. The profiles of the corresponding pulses are depicted in Fig.~\ref{f2}(b). Here $\Gamma=0.04$, $\gamma=4$, $T=10$. Fig.~\ref{f2}(a) shows that the fidelity evolutions under these pulses are almost indistinguishable, i.e. both the optimal pulses have much similar performances as the ideal pulses. In the inset of Fig.~\ref{f2}(a) we plot the convergence behaviors of the two algorithms. We choose the maximum of iteration times $k_{max}=6000$ for both algorithms and the learning rate $\alpha=10,1$ for SGD and Adam respectively. Evidently, in this case Adam converges faster: the same fidelity values are obtained after about 3000 iteration times for SGD but less than 500 times for Adam. Adam has the advantage to converge far faster than SGD after several improvements. Hence from now on, we employ Adam alone. Fig.~\ref{f2}(b) shows the corresponding profiles of control pulses used in Fig.~\ref{f2}(a). We can see that the differences are subtle, especially for Adam and SGD.

	In the above search, we only use a single sinusoidal control function and the fidelity improvement is small. To achieve a more significant fidelity improvement, now we propose a combinational control function
	\begin{equation}
		c(t)=\sum_{i=0}^{M-1}I_{i}\sin\left[\left(i+1\right)\omega t\right],
		\label{combinational}
	\end{equation}
	which is a combination of Fourier sinusoidal components and also satisfies the zero-area pulse conditions. Here $M$ indicates the number of Fourier components and we consider $M=10$ in this work.
	
	In Fig.~\ref{f3}(a), we plot the fidelity $F$ as a function of the rescaled time $t/T_{tot}$ under the ideal closed-system pulses and the two kinds of optimal open-system pulses. The profiles of the corresponding pulses are depicted in Fig.~\ref{f3}(b). Here we still take $\Gamma=0.04$, $\gamma=4$, and $T=10$. For the combinational pulses, we set the initial pulse intensity sequence $\textbf{\textit{I}}^{i}=[10,10,\cdots ,10]$, the maximum of iteration times $k_{max}=1000$ and the learning rate $\alpha=1$. Obviously, the single pulses in Eq.~(\ref{pulse2}) are inferior to the combinational ones, which certainly also outweigh the ideal closed-system pulses. $F(T_{tot})=0.949$ for the combinational pulses while $F(T_{tot})=0.934$ for the single pulses. The reason may be that the combinational control function can be updated more elaborately in each iteration.

\begin{figure}[!htbp]
\centering
\includegraphics[width=\columnwidth]{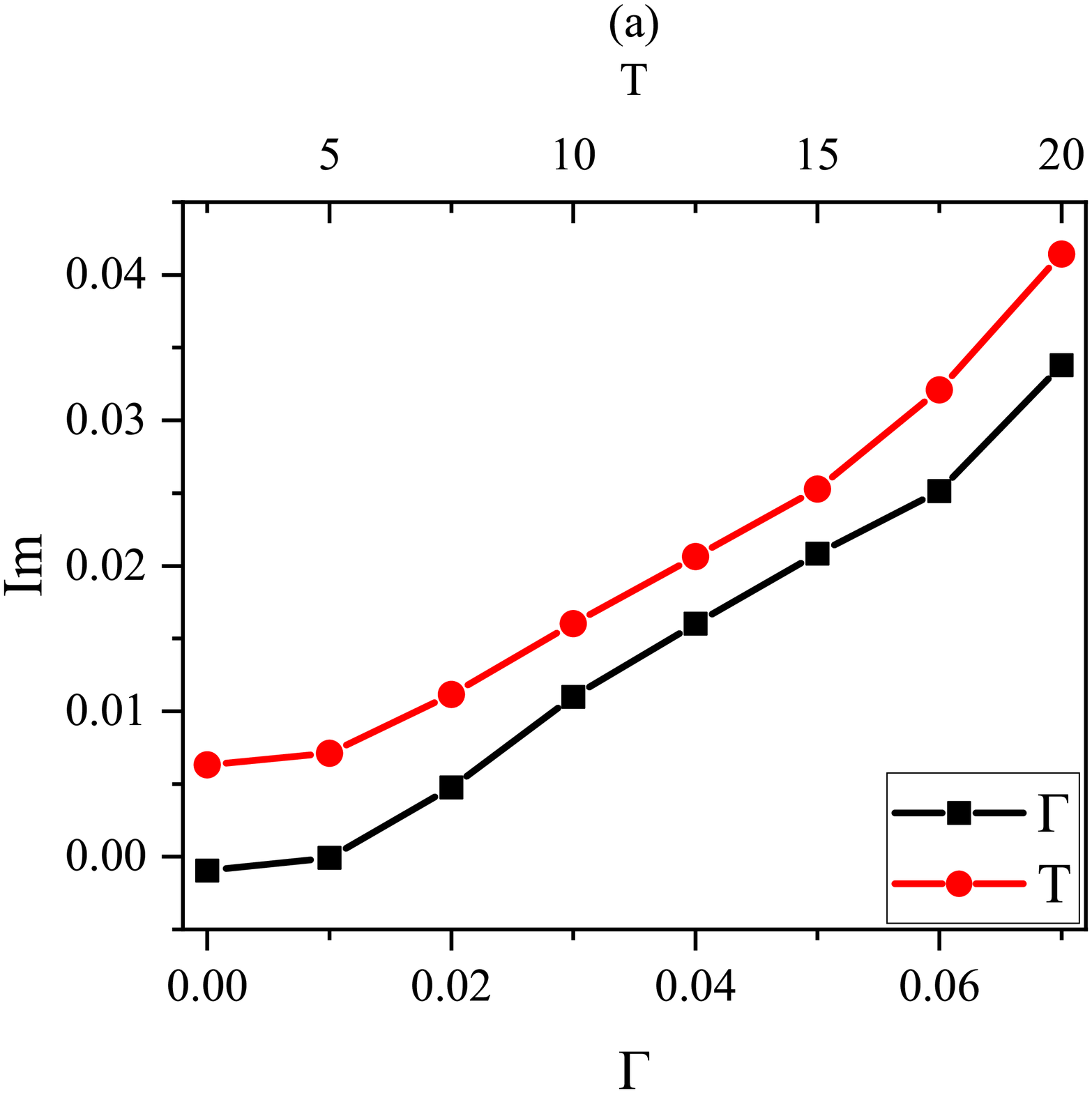}
\includegraphics[width=\columnwidth]{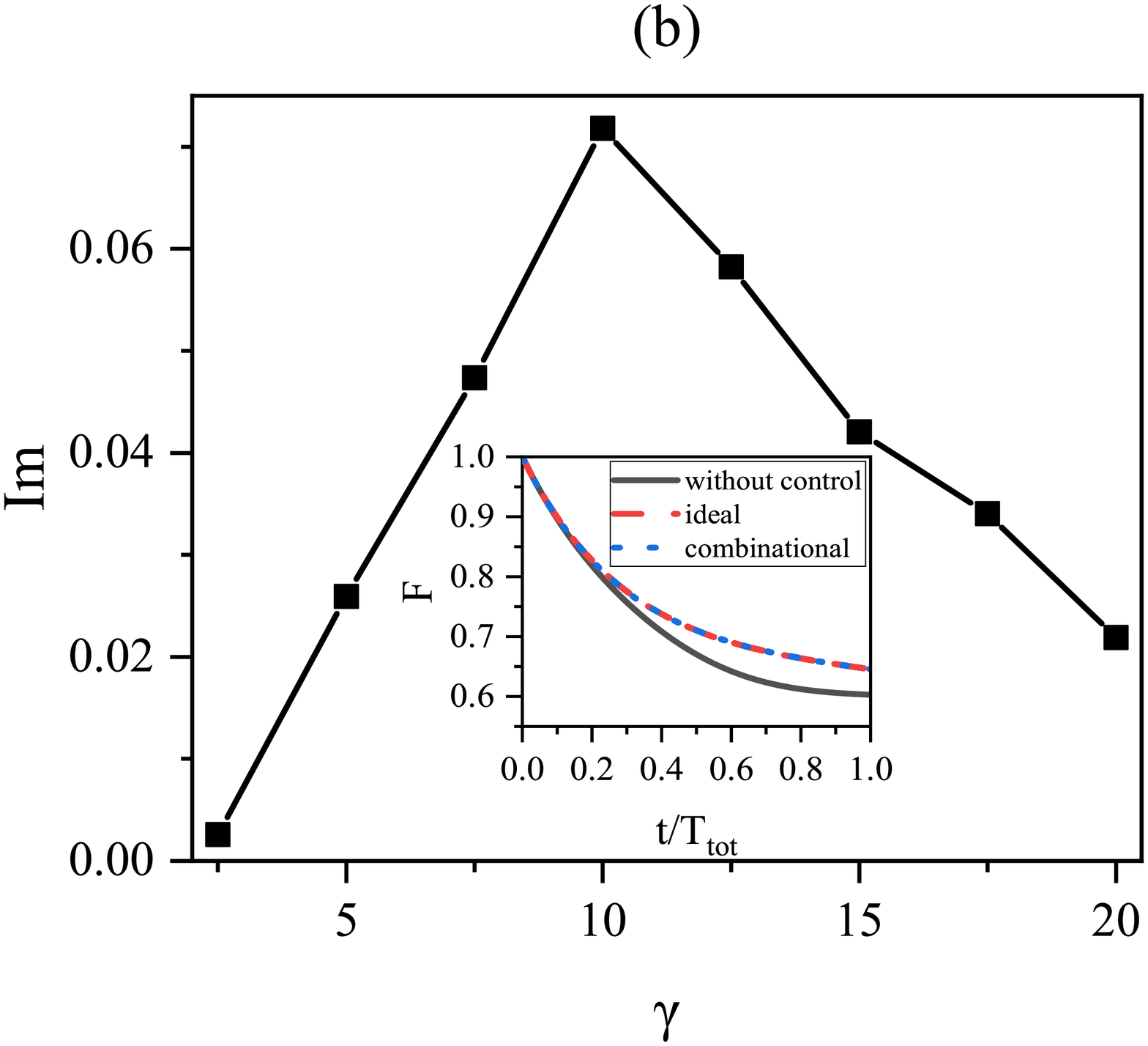}
\caption{(Color on line) The fidelity improvement $Im$ for different environmental parameters. Other parameters are the same as in Fig.~\ref{f3}(a). (a) $\Gamma$, $\gamma=4$, $T=10$ and $T$, $\Gamma=0.04$, $\gamma=4$. (b) $\gamma$, $\Gamma=0.04$, $T=10$.}
\label{f4}
\end{figure}

	In order to visually show the above fidelity enhancement, we define the fidelity improvement $Im$:
		\begin{equation}
			Im=F(T_{tot})^{combi}-F(T_{tot})^{ideal}.
		\end{equation}
	Here $F(T_{tot})^{combi}$ and $F(T_{tot})^{ideal}$ are the final fidelities obtained by combinational pulses and ideal pulses, respectively. 
	
	Now we discuss the influences of different environmental parameters $\Gamma$, $\gamma$ and $T$ on the final fidelity improvement $Im$. In Fig.~\ref{f4}(a) we plot the fidelity improvement $Im$ for different $\Gamma$ and $T$. Here for different $T$, we take $\Gamma=0.04$ and $\gamma=4$. For different $\Gamma$, we take $\gamma=4$ and $T=10$. Obviously, $Im$ grows with the increasement of parameter $\Gamma$ or $T$. That is to say, a more stronger bath (higher $T$ or $\Gamma$) provides more room for the SSPs to boost the fidelity. Fig.~\ref{f4}(b) plots the effects of the parameter $\gamma$ on the fidelity improvement $Im$. $\gamma$ represents the memory time of the environment and larger $\gamma$ corresponds to a more memoryless environment. In Fig.~\ref{f4}(b) $Im$ first increases and then decreases with increasing $\gamma$. There is a peak around $\gamma=10$. Since a more Markovian bath affects the system more severely, $Im$ should increase monotonically with increasing $\gamma$, as shown in Fig.~\ref{f4}(a). However, the effects of the control also become weak with increasing $\gamma$ \cite{wang2021,nonmarkov1,nonmarkov2}. Physically, the information from the system to the environment loses faster for a larger $\gamma$, and as a result only little time is given to control the state of system. This observation is in accordance with Refs.~\cite{nonmarkov1,nonmarkov2}, which shows that the effectiveness of control can be boosted when the environment has a longer memory time. When $\gamma<10$, the first effect dominates while the second one dominates when $\gamma>10$. There is a critical $\gamma$ which corresponds to the largest $Im$ for certain parameters $\Gamma$ and $T$. In the Markovian limit, the dynamics is given by the Lindblad equation in Eq.~(\ref{markov}). In the inset of Fig.~\ref{f4}(b) we plot the fidelity $F$ versus the rescaled time $t/T_{tot}$ in a Markovian environment. Here $\Gamma=0.04$ and $T=10$. Clearly, the control loses its effectiveness: there is only a small improvement compared with the free evolution case. It is also worth noting that the combinational schemes lose their advantage: the two control curves are almost indistinguishable.
	
\begin{figure}[!htbp]
\centering
\includegraphics[width=\columnwidth]{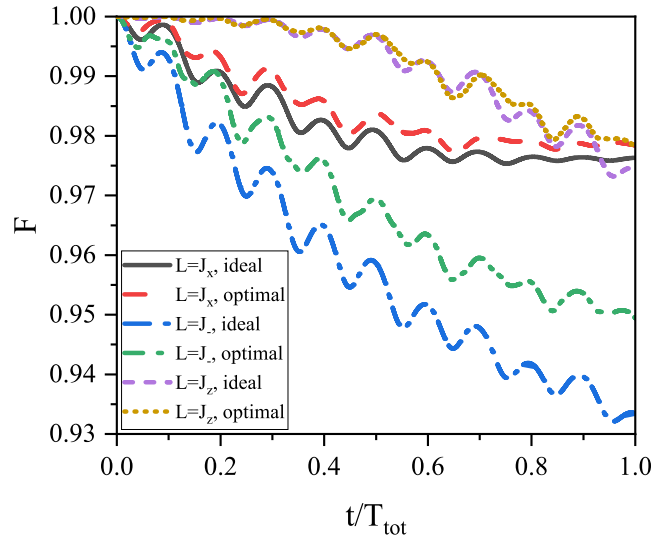}
\caption{(Color on line) The fidelity $F$ versus the rescaled time $t/T_{tot}$ with the help of the ideal closed-system pulses and optimal (combinational) open-system pulses when $L=J_{x}$, $J_{-}$ or $J_{z}$. Here $\Gamma=0.04$, $\gamma=4$ and $T=10$.}
\label{f5}
\end{figure}

	In above discussion, we only consider $L=J_{-}$. Are the above schemes still effective for different $L$? Next we consider $L=J_x,J_z$. Fig.~\ref{f5} shows that the fidelity improvement can be realized for all three cases. This is to say, the control schemes are still effective. However, $Im$ is different for same environmental parameters $\Gamma=0.04$, $\gamma=4$ and $T=10$. It shows that $Im$ is the most significant for the case $L = J_-$, and $L=J_x$ is the next while $L=J_z$ is the last.

\section{conclusions}
	High-accuracy quantum control is required for the realization of adiabatic speedup, especially when the system is immersed in a relatively strong environment. The ideal pulses have been theoretically derived by P-Q partitioning technique for closed systems. To address the open-system cases, in this work we apply the SSPs to search the optimal pulses for significant adiabatic speedup, which have the advantage that the detrimental effects of the environment can be combated. Specifically, we consider a qutrit system in a non-Markovian finite-temperature environment and use SGD and Adam algorithms to design the optimal pulses. For the single pulses, we find that the fidelity obtained by SSPs increases slightly compared with the ideal case. We then construct the combinational control function to obtain a more significant adiabatic speedup. We define a fidelity improvement $Im$ to demonstrate the advantage of our SSPs. $Im$ is significant for a relatively strong environment, which indicates that the SSPs are more powerful when the bath affects the system more severely. However, as the controllability becomes weak for a more Markovian environment, $Im$ decreases with increasing $\gamma$ for large $\gamma$. The maximum of improvement $Im$ can be obtained for certain environmental parameters. Our investigation demonstrates that the SSPs are powerful tools for the optimal pulse design in open systems.

\begin{acknowledgements}
	We would like to thank Jing Wu for helpful discussions. This paper is based upon work supported by National Natural Science Foundation of China (Grant No. 11475160) and the Natural Science Foundation of Shandong Province (Grants No. ZR2021LLZ004 and No. ZR2014AM023).
\end{acknowledgements}

\end{document}